\documentclass[preprint, 12pt]{elsarticle}

\usepackage{graphicx}
\usepackage{balance}
\usepackage{todonotes}
\usepackage{multirow}
\usepackage{algorithm}
\usepackage{algpseudocode}
\usepackage{pgfplots}
\usepackage{amsmath}
\usepackage{amsthm}
\pgfplotsset{compat=1.18}
\newtheorem{theorem}{Theorem}
\usepackage{mathtools}

\algnewcommand\algorithmicassert{\texttt{assert}}
\algnewcommand\Assert[1]{\State \algorithmicassert(#1)}
\theoremstyle{definition}
\newtheorem{definition}{Definition}[section]
\DeclarePairedDelimiter\abs{\lvert}{\rvert}

\author[udc]{David Soler\corref{cor1}}
\author[udc]{Carlos Dafonte}
\author[uvigo]{Manuel Fernández-Veiga}
\author[uvigo]{Ana Fernández Vilas}
\author[udc]{Francisco J. Nóvoa}
\cortext[cor1]{Corresponding author. Contact: david.soler@udc.es}

\affiliation[udc]{organization={CITIC, Universidade da Coruña},
            city={A Coruña},
            country={Spain}}
\affiliation[uvigo]{organization={atlanTTic, Universidade de Vigo},
            city={Vigo},
            country={Spain}}

\journal{Computer Networks}

\title{A Privacy-preserving key transmission protocol to distribute QRNG keys using zk-SNARKs}

\begin{document}
\begin{frontmatter}

\begin{abstract}
High-entropy random numbers are an essential part of cryptography, and Quantum Random Number Generators (QRNG) are an emergent technology that can provide high-quality keys for cryptographic algorithms but unfortunately are currently difficult to access. Existing Entropy-as-a-Service solutions require users to trust the central authority distributing the key material, which is not desirable in a high-privacy environment. In this paper, we present a novel key transmission protocol that allows users to obtain cryptographic material generated by a QRNG in such a way that the server is unable to identify which user is receiving each key. This is achieved with the inclusion of Zero Knowledge Succinct Non-interactive Arguments of Knowledge (zk-SNARK), a cryptographic primitive that allow users to prove knowledge of some value without needing to reveal it. The security analysis of the protocol proves that it satisfies the properties of Anonymity, Unforgeability and Confidentiality, as defined in this document. We also provide an implementation of the protocol demonstrating its functionality and performance, using NFC as the transmission channel for the QRNG key.
\end{abstract}

\begin{highlights}
\item We propose a key distribution protocol to provide access to QRNG entropy to users.
\item Our protocol employs zk-SNARKs to prevent the server from knowing who is requesting the key.
\item We provide an implementation of the protocol that distributes the keys over NFC.
\end{highlights}


\end{frontmatter}

\section{Introduction} 
\label{sec:introduction}

Random numbers have always been a staple in cryptography. Every cryptographic algorithm, from symmetric encryption like AES to asymmetric encryption like RSA, requires random numbers as key material or initialisation parameters. The most common sources of randomness for cryptographic purposes are Pseudo-Random Number Generators (PRNG), which are deterministic processes that are initialised from a specific seed. Furthermore, PRNGs can contain vulnerabilities in their implementations~\cite{rng1, rng2} that make them vulnerable to attacks. Thus, PRNGs usually produce low-entropy key material, which can compromise the entire cryptographic scheme: no matter how theoretically secure an encryption algorithm is, it could be easily cracked if an attacker manages to obtain the secret key by exploiting its insecure generation. This is not only a theoretical issue: low-entropy key generation has been used as an entrypoint to break confidentiality in well-known cryptosystems~\cite{low-entropy1, low_entropy2, low-entropy3}.

Quantum Random Number Generators (QRNG)~\cite{qrng2, qrng3, qrng4} are a different type of Random Number Generators that make use of the intrinsically unpredictable behaviour of quantum particles as their source of entropy. Because of this, the key material they generate is of higher quality than that generated by a PRNG, and its use could enhance the security of current cryptographic algorithms. Unfortunately, the technology is not yet fully developed and QRNGs are an expensive asset, so  these devices cannot be widely distributed and end users have very limited access to them. In a context that may require a high level of privacy, such as private communications, users will benefit from having access to cryptographic keys with high entropy, such that it is harder for attackers to guess the key they have used. Furthermore, providing easy access to this key material could encourage users to take their privacy seriously and improve their overall communication security.

The objective of this work is to provide a method for end users to obtain higher-quality cryptographic material by defining a key transmission protocol in which a server with access to the QRNG distributes random keys in a secure manner to users that request them. As part of the process, the server will require clients to authenticate themselves before providing the QRNG key material.

However, existing Entropy-as-a-Service solutions \cite{eaas} require users to trust its key distribution server, since this server can trace which keys it has provided to each user. A malicious or compromised server could use the keys it has transmitted to decrypt users' communications or even impersonate them. Indeed, the requirement to trust the server is a common concern in key distribution protocols. This work aims to provide a solution to that problem, designing a key transmission protocol that seeks a balance between requiring authentication (i.e., requesting private information from users so that access is restricted to those who can prove their identity) and protecting the users' privacy. This requirement is achieved with the inclusion of Zero Knowledge Succinct Non-interactive Arguments of Knowledge (zk-SNARK) as proofs of authentication, a cryptographic primitive that allows users to prove that they know some value without needing to reveal it, such that any verifier could be convinced of that fact. 

Our protocol is divided in two main phases: in the first phase, a user performs authentication and in the second phase the user requests the QRNG key by presenting a zk-SNARK proof. The inclusion of zk-SNARKs allows us to clearly decouple the two phases, such that presenting the zk-SNARK proof in the second phase does not reveal any information (not even to the server) about the credentials used in the first phase. Thus, users can prove that they have previously authenticated themselves, in such a way that the server could not distinguish the user from any other user that has ever performed authentication. This prevents the key distribution server from associating a key to a specific user, which would allow the server to decrypt any message that uses this key. Our protocol also provides protection against replay attacks: if an attacker tries to reuse another user's zk-SNARK proof, the server will detect the attempt and abort the execution.

Since random numbers are required in all widely used cryptographic algorithms, the QRNG-generated values distributed by the protocol could be useful in a wide range of applications. For example, scenarios where users need to establish communication channels with a high level of confidentiality could employ the random numbers as high-quality key material for symmetric encryption. Another application is in multifactor authentication: possession of a random number generated by this protocol also proves that its owner has successfully performed authentication, therefore a server could require users to execute this protocol and present the QRNG-generated value to provide access.

In summary, our work presents the following contributions:
\begin{enumerate}
    \item A Key Transmission Protocol for the distribution of QRNG keys. The protocol requires two distinct interactions between clients and server: one for authentication and another for receiving the QRNG key.
    \item A privacy-preserving method of authentication for the key Transmission Protocol. The authentication is defined in such a way that it is impossible for the server to track who is receiving each key, thus preventing the server from spying on clients' communications that use these keys. This allows users to execute the protocol even if they do not trust the server generating the keys. The protocol meets the security requirements of Anonymity, Unforgeability and Confidentiality, as defined in this document. We remark that the Unforgeability property also prevents replay attacks, in which an attacker tries to present another user's proof as its own.
    \item An implementation of the proposed Key Transmission Protocol. The implementation simulates a real use case, in which the server is located in some specific location users have to attend to. Part of the communication is performed through NFC and the key is stored in the users' mobile phone. Two different zk-SNARK libraries are employed for proof generation and validation, and the server uses two QRNGs for key generation. The time complexity of the protocol is estimated, measuring the proof generation time for different parameters, the NFC transmission time and the key generation rate for both QRNGs.
\end{enumerate}

The rest of the document is organised as follows: Section~\ref{sec:literature} will review related works and compare them with our contribution. Section~\ref{sec:tech} will introduce the required technical background to the reader. Section~\ref{sec:protocol} explains in detail all steps involved in the execution of the protocol, as well as the entities that participate in it. Section~\ref{sec:security} discusses the security properties of the protocol. Section~\ref{sec:impl} analyses the proof-of-concept implementation of all entities involved in the protocol, and in  Section~\ref{sec:test} the performance of the implemented applications is tested. Finally, Section~\ref{sec:conclusion} will conclude this document and discuss future improvements.     

\section{Related Work}
\label{sec:literature}

Many of the recent improvements in sources of entropy have focused on Distributed Randomness Beacons (DRB) \cite{pub-rand, drb}. In these schemes, a group of users can generate a public random value that can be verified by all participants and other parties, ensuring that no user could have introduced bias in the calculation. While verifiable public randomness is useful in multiple scenarios, there is still the need for sources of entropy that distribute the generated random numbers in a private manner.

There exist multiple services that provide high entropy randomness on-demand, such as \textit{random.org} \cite{rand-org}. NIST has designed an architecture for the generation and distribution of random numbers \cite{eaas}, which requires clients to present an authentication token. In these schemes the server knows which client is receiving each random values, and thus any cryptographic material that is generated through them. This represents a privacy concern for clients, which may not trust the server that provides the source of entropy.

The goal of authentication usually comes into conflict with the privacy of the users: the authentication server may require more information than the client is willing to give. Zero Knowledge Proofs are a family of methods by which a prover can prove to a verifier that he knows some private information without needing to reveal it. Their use in authentication protocols allow users to prove their identity without revealing sensitive information about themselves. In Non-interactive variants of Zero Knowledge Proofs, the prover and the verifier do not need to directly establish communication: instead, the prover publishes a proof that can be verified later by any other entity.

Non-interactive Zero-Knowledge authentication~\cite{nizk-auth} is achieved in~\cite{nizk-auth1} by requiring authorised nodes to solve a hard problem (in this case, graph isomorphism), which users from outside the network would not be able to solve. Likewise,~\cite{nizk-auth2} also uses graph isomorphism as authentication mechanism, using a Merkle Tree as a commitment to turn the authentication into non-interactive. Both of these protocols assume a list of clients capable of solving this hard problem, making it difficult to add new authorised clients to the system.

The authors of~\cite{bp-akaa} define a scheme in which users belonging to certain trust domains can authenticate each other and establish communication sessions. A method for registering new users is defined, but the defence against replay attacks is weak, since it uses timestamps: it does not protect against replays that happen in short span of time.

The scheme defined in~\cite{elearn} allows educational institutions to submit a learner's diploma into a public blockchain. A custom zero-knowledge proof is employed in multiple occasions, such as protecting the learner's identity or proving the grades of the learner. The protocol does not provide any direct protection against replay attacks, but replaying a proof is not sufficient to impersonate a legitimate learner: an attacker would also need to know that learner's secret key. 

Among NIZK algorithms, zk-SNARKs are a popular alternative that can also be used in authentication protocols. The authors of~\cite{nizk-auth3} propose an authentication protocol for healthcare environments in which a set of pre-registered clients authenticate themselves with a zk-SNARK validated with a Smart Contract with the objective of establishing a secure communication channel. In~\cite{sims}, a certificate authority is inserted into the scheme to validate users' parameters, which they can later provide proof of. However, it presents no defence against attackers that could intercept another user's proof and later present it as their own. In~\cite{eoltaa}, the focus is defining a system that can provide Linkability and Traceability to identify malicious users, while protecting the privacy of honest users. Since the authors orient their work to a voting application, the list of authorised users is predefined, and cannot be changed. 

The concept of Commitments and Nullifiers, which are extensively used in this work, are introduced in \cite{zcash}. As in the proposed protocol, in ZCash users can create a Commitment when they earn the right to perform an action and they must publish a Nullifier to consume it, accompanied by a zk-SNARK verifying the relationship between these hashes.

Table~\ref{table:comparison} compares the protocol we propose with the works mentioned in this Section. As shown, our protocol is capable of introducing new users dynamically into the system, does not require external infrastructure outside the control of the scheme, and provides protection against replay attacks.

\begin{table*}
    \centering
    \resizebox{\textwidth}{!}{
    \begin{tabular}{cccccc}
    \hline
    Scheme            & Purpose                 & Authentication Method   & Client Registration & External Infrastructure & Replay protection \\ \hline
    \cite{nizk-auth1} & Authentication          & Isomorphism             & No              & No                      & No                \\
    \cite{nizk-auth2} & Authentication          & Isomorphism             & No              & No                      & No                \\
    \cite{bp-akaa}    & Key Agreement           & Discrete Logarithm      & Yes             & No                      & Timestamp         \\
    \cite{elearn} & E-learning records           & Zero Knowledge Proof   & Yes             & Yes (Blockchain)        & Secret key        \\
    \cite{eoltaa}     & Voting                  & zk-SNARK                & No              & Yes (Blockchain)        & Yes               \\
    \cite{sims}       & Identity Management     & zk-SNARK                & Yes             & Yes (Blockchain)        & No                \\
    \cite{nizk-auth3} & Key Agreement           & zk-SNARK                & No              & Yes (Ethereum)          & Yes               \\
    Ours              & Key Transmission        & zk-SNARK/Merkle Tree    & Yes             & No                      & Yes               \\ \hline
    \end{tabular}}
    \caption{Comparison between other works that use non-interactive zero-knowledge authentication and our proposed protocol.}
    \label{table:comparison}
\end{table*}

\section{Background} 
\label{sec:tech}

To achieve our privacy-preserving Key Transmission Protocol, we will need to decouple the Authentication and Key Request steps, such that the server will not be able to link clients performing the latter with the credentials they used in the former. To that end, we employ the cryptographic primitive zk-SNARK, which is formally defined in Section \ref{sec:snark}.

We use zk-SNARKs in this work to prove membership to a set of valid users. To speed up zk-SNARK proof generation, the aforementioned set is structured as a Merkle Tree, in which proving membership only takes $log_2(N)$ steps where $N$ is the total number of users. Finally, we employ assymetric cryptography to securely transmit the QRNG key to the user in the last step of the protocol.

\subsection{zk-SNARKs}
\label{sec:snark}

Zero Knowledge Succinct Non-Interactive Arguments of Knowledge (zk-SNARKs) are a subset of non-interactive zero-knowledge proofs. A zk-SNARK is built from a relation $R$ between a public Statement $x$ and a private Witness $w$. If the relation between $w$ and $x$ holds, then $(x, w) \in R$. A security parameter $\lambda$ can be derived from the description of $R$. Formally, a zk-SNARK scheme $\Pi$ for a relation $R$ is composed of the triplet $(\mathsf{Gen}, \mathsf{Prove}, \mathsf{Verify})$, defined as follows:

\begin{itemize}
    \item $\mathsf{Setup}(R) \rightarrow (pk, vk)$: Takes a relation and outputs the Common Reference Key, which is divided into the Proving Key $pk$ and the Verification Key $vk$.
    \item $\mathsf{Prove}(pk, x, w) \rightarrow \pi$: From $pk$, a witness $w$ and a statement $x$, a proof $\pi$ is generated. 
    \item $\mathsf{Verify}(vk, x, \pi) \rightarrow Accept/Reject$: From $vk$, a proof $\pi$ and the corresponding statement $x$, outputs $Accept$ or $Reject$. In the context of this work, a \textit{valid proof} refers to any $(x_v, \pi_v)$ such that $\mathsf{Verify}(vk, x_v, \pi_v) = Accept$.
\end{itemize}

A zk-SNARK scheme must also satisfy the following requirements~\cite{redacting}:
\begin{itemize}
    \item Completeness. The probability 

\[
\operatorname{Pr}
\left[
\begin{array}{c}
(pk,vk) \gets \mathsf{Gen}(R) \\
\pi \gets \mathsf{Prove}(pk, x, w) \\
(x, w) \in R
\end{array}
:
\begin{array}{c}
\mathsf{Verify}(vk, x, \pi) = Reject
\end{array}
\right]
\]

is negligible. Intuitively, this means that an honest prover is able to convince a verifier that $(x, w) \in R$.
    \item Knowledge Soundness. For every efficient adversary $A$, there exists an efficient extractor $Ext_A$ with access to the internal state of $A$ such that the probability 

\[
\operatorname{Pr}
\left[
\begin{array}{c}
((pk,vk), aux) \gets \mathsf{Gen}(R) \\
(x, \pi) \gets \mathsf{A}(R, aux, (pk, vk)) \\
w \gets Ext_A(R, aux, (pk, vk))
\end{array}
:
\begin{array}{c}
(x, w) \notin R \\
\land \mathsf{Verify}(vk, x, \pi) = Accept
\end{array}
\right]
\]
    
    is negligible, where $aux$ is an auxiliary input produced by $\mathsf{Gen}$. Intuitively, this means that dishonest provers could not generate a valid proof if they do not know $w$.
    \item Zero-knowledge. For every adversary A acting as a black box and $(x, w) \in R$, there exists a simulator $\mathsf{Sim}((pk, vk), aux, x)$ such that the following equality holds:

\[
\operatorname{Pr}
\left[
\begin{array}{c}
((pk,vk), aux) \gets \mathsf{Gen}(R) \\
\pi \gets \mathsf{Prove}(pk, x, w) \\
\end{array}
:
\begin{array}{c}
\mathsf{A}((pk, vk), aux, x, \pi) = 1
\end{array}
\right]
\]
$\approx$
\[
\operatorname{Pr}
\left[
\begin{array}{c}
((pk,vk), aux) \gets \mathsf{Gen}(R) \\
\pi \gets \mathsf{Sim}((pk, vk), aux, x) \\
\end{array}
:
\begin{array}{c}
\mathsf{A}((pk, vk), aux, x, \pi) = 1
\end{array}
\right]
\]

where $aux$ is an auxiliary input produced by $\mathsf{Gen}$. Intuitively, this means that an attacker cannot find out anything about a witness $w$ from a proof $\pi$, a statement $x$ and a key pair $(pk, vk)$.
    \item Weak Simulation-Extractability. For every efficient adversary $A$ with access to an oracle $O$, there exists an efficient extractor $Ext_A$ with access to the internal state of $A$ such that the probability 

\[
\operatorname{Pr}
\left[
\begin{array}{c}
((pk,vk), aux) \gets \mathsf{Gen}(R) \\
(x, \pi) \gets A^{O(aux)}(R, aux, (pk, vk)) \\
w \gets Ext_{A}(R, aux, (pk, vk))
\end{array}
:
\begin{array}{c}
x \notin Q \\ 
\land (x, w) \notin R \\ 
\land \mathsf{Verify}(vk, x, \pi) = Accept

\end{array}
\right]
\]
    is negligible, where $aux$ is an auxiliary input produced by $\mathsf{Gen}$ and Q is the list of $x_i$ requested to the oracle. This means that $A$ cannot generate a valid proof for a statement that was not obtained from the oracle. Note that there is a stronger notion of Simulation-Extractability that requires both the statement $x$ and the proof $\pi$ not to be included in $Q$. However, for this work we only require the weak version of Simulation-Extractability.
    \item Succinctness. The proof size is polynomially bounded by $\lambda$. The execution time of $\mathsf{Verify}$ is bounded linearly by the size of $x$ and polynomially by $\lambda$.
\end{itemize}

In practice, zk-SNARK schemes can be built by different cryptographic techniques. The most popular designs, including Groth16 \cite{groth16} (which we will use in this work) involves the transformation of a computation into an arithmetic circuit. The performance of a zk-SNARK scheme, specially during setup and proof generation, heavily depends on the number of constraints of this circuit \cite{zk-time}, which are determined by the computation that is to be verified. 

\subsection{Merkle Tree}

Merkle Trees~\cite{merkle-original} are a data structure in the form of a binary tree. Each element is identified as $h_{i,j}$, where $i$ is the level of the tree and $j$ the position inside said level. A Merkle Tree of depth $n$ contains at most $l = 2^n$ leaf nodes $h_{n,j}$ containing a hash, and intermediate nodes take the value $h_{k, j} = H(h_{k+1, m} \mathbin\Vert h_{k+1, m+1})$, where $\mathbin\Vert$ is the concatenation operator, $H$ is a hash function and $m = 2 j$. The node $h_{0,0}$ is called $root$, and its value is influenced by every other node in the Tree.

The main advantage of Merkle Trees is that checking if an element $e$ exists inside the list of leaf nodes is only $\mathcal{O}(\log l)$, while the same operation for a list would be $\mathcal{O}(l)$. This is achieved by providing a validation list $\{h_{n-1, k_{n-1}}, h_{n-2, k_{n-2}}, ..., h_{1, k_1}\}$ and the root $h_{0, 0}$ of the Merkle Tree. If it is possible to reconstruct the provided root from $H(e)$ and the validation list, then $e$ is proven to be inside the Merkle Tree. This method requires the hash function $H$ used to construct the Merkle Tree to be collision-resistant~\cite{merkle}: otherwise, it could be possible for malicious users to prove that a fake element $e_f$ is inside the Merkle Tree when it is not.  

For the scope of this work, the following operations are defined for Merkle Trees:

\begin{enumerate}
    \item $\mathsf{EmptyTree}(n) \rightarrow T$: Initialises an empty Merkle Tree of depth $n$, with all of its leaf nodes set to $H(0)$.
    \item $\mathsf{AddLeaf}(T, h) \rightarrow T'$: Takes a Merkle Tree $T$ and a hash $h$ and returns a modified tree, in which the leftmost $H(0)$ is substituted by $h$. The intermediate and root nodes are also updated accordingly.
    \item $\mathsf{GetRoot}(T) \rightarrow R$: Takes a Merkle Tree $T$ and returns the root $h_{0,0}$.
    \item $\mathsf{GetIndexOf}(T, h) \rightarrow index$: Takes a Merkle Tree $T$ and a hash $h$ and returns $i$ such that $h_{n,i} = h$ for some $h_{n,i}$ in the Tree.
    \item $\mathsf{ValidationList}(T, i) \rightarrow Val = \{h_{n-1, k_{n-1}}, h_{n-2, k_{n-2}}, ..., h_{1, k_1}\}$: Takes a Merkle Tree $T$ and an index $i$ and returns the validation list required to validate the leaf $h_{n,i}$.
    \item $\mathsf{IsLeafOfTree}(h, i, Val, R) \rightarrow true/false$: Takes a hash $h$, its index $i$, a validation list $Val = \{h_{n-1, k_{n-1}}, h_{n-2, k_{n-2}}, ..., h_{1, k_1}\}$ and a root $R$ and outputs $true$ if $h$ is the leaf with index $i$ of the tree with root = $R$, and $false$ otherwise.
\end{enumerate}

\subsection{Key Encapsulation Mechanism}

A Key Encapsulation Mechanism (KEM) \cite{kem, rfc9180} is a method for distributing key material between two parties employing assymetric cryptography. The sender uses the receiver's public key to encrypt some input key material such that only the receiver can decrypt it. Both parties must agree on the assymetric encryption algorithm that will be used for the encapsulation. In Hybrid Public Key Encryption \cite{hpke}, the key material is later passed through a Key Derivation Function to obtain a symmetric key. 

A KEM is composed by the following operations:

\begin{itemize}
    \item $\mathsf{KeyGen}(\lambda) \rightarrow (pk, sk)$: Takes a security parameter $\lambda$ and generates a symmetric key pair.
    \item $\mathsf{Encap}(ikm, pk) \rightarrow c$: Encrypts some input key material $ikm$ with a public key $pk$. 
    \item $\mathsf{Decap}(c, sk) \rightarrow km$: Decrypts a ciphertext with a secret key $sk$ to obtain the key material.
\end{itemize}

\section{Privacy-Preserving Key Transmission Protocol}
\label{sec:protocol}

\subsection{Outline}
\label{sec:outline}
The protocol involves two different interactions between Users and the Server: the Authentication and the Key Request. The main contribution of our protocol is the decoupling between the authentication and the key request, which is achieved through the use of zk-SNARKs: while requesting a key, Users must prove that they have previously performed authentication.

During the Authentication, Users must present valid credentials that certify their identity, such as a digital certificate. If those credentials are correct, they are allowed to publish a \textit{Commitment}, which the Server includes in a structure of valid Commitments (which, for efficiency reasons, is a Merkle Tree). Thus, a Commitment identifies a specific User, because is published alongside authentication information. It also represents the User's right to obtain a QRNG key in later steps.

Users must generate a \textit{Nullifier}, which is a value that is cryptographically related to their Commitment: they are generated from the same inputs, but only the User that has created both knows that they are related. They also generate a zk-SNARK proof that certifies the relationship between the Commitment and Nullifier, but without revealing the former. The proof also certifies that the Commitment is inside the structure of valid Commitments. 

To perform the Key Request, Users send the aforementioned Nullifier and zk-SNARK proof to the Server, which validates those values. If the proof is valid, the Server (1) knows that there exists a Commitment related to this specific Nullifier inside the structure of valid Commitments, but (2) does not know which of all valid Commitments it is. We remark that (1) proves that the User has performed the Authentication exchange, while (2) ensures that neither the Server or any other entity can know the identity of the User. Finally, the Server stores the Nullifier to prevent replay attacks: any other Key Request that contains the same Nullifier will be rejected.

\subsection{Architecture}

In this protocol, there are three main entities:
\begin{itemize}
    \item \textit{User (U)}: executes the protocol and starts the communication for both Steps. In the Authentication Step, Users must provide their certificate and generate a Commitment. Then, they generate a Nullifier and a zk-SNARK linking it to the Commitment.
    \item \textit{Authentication Server} (AS): receives authentication requests from Users. When a User's certificate is successfully validated, the AS inserts the provided Commitment into the Commitment Merkle Tree and updates the required nodes and root. The AS must also provide the Merkle Tree to anyone that requests it.
    \item \textit{Proof Validation Server} (PVS): receives zk-SNARKs from Users and validates them. The PVS stores the list of all published Nullifiers, so it rejects any proof that is accompanied by a Nullifier that has already been published. The PVS requires information about the Merkle Tree to validate zk-SNARKs, so it must communicate with the AS to obtain this data structure. It is the only entity in the protocol that can access the QRNG, which it uses to generate key material to Users that present valid zk-SNARKs.
\end{itemize}

The AS and PVS are logical entities, so they can be implemented either as one or two different programs. All information hosted by the AS and the PVS (namely, the Commitment Merkle Tree and the Nullifier List) is exposed so that Users can freely access it without compromising the security of the protocol.

\begin{figure}[H]
    \centering
    \includegraphics[width=\columnwidth]{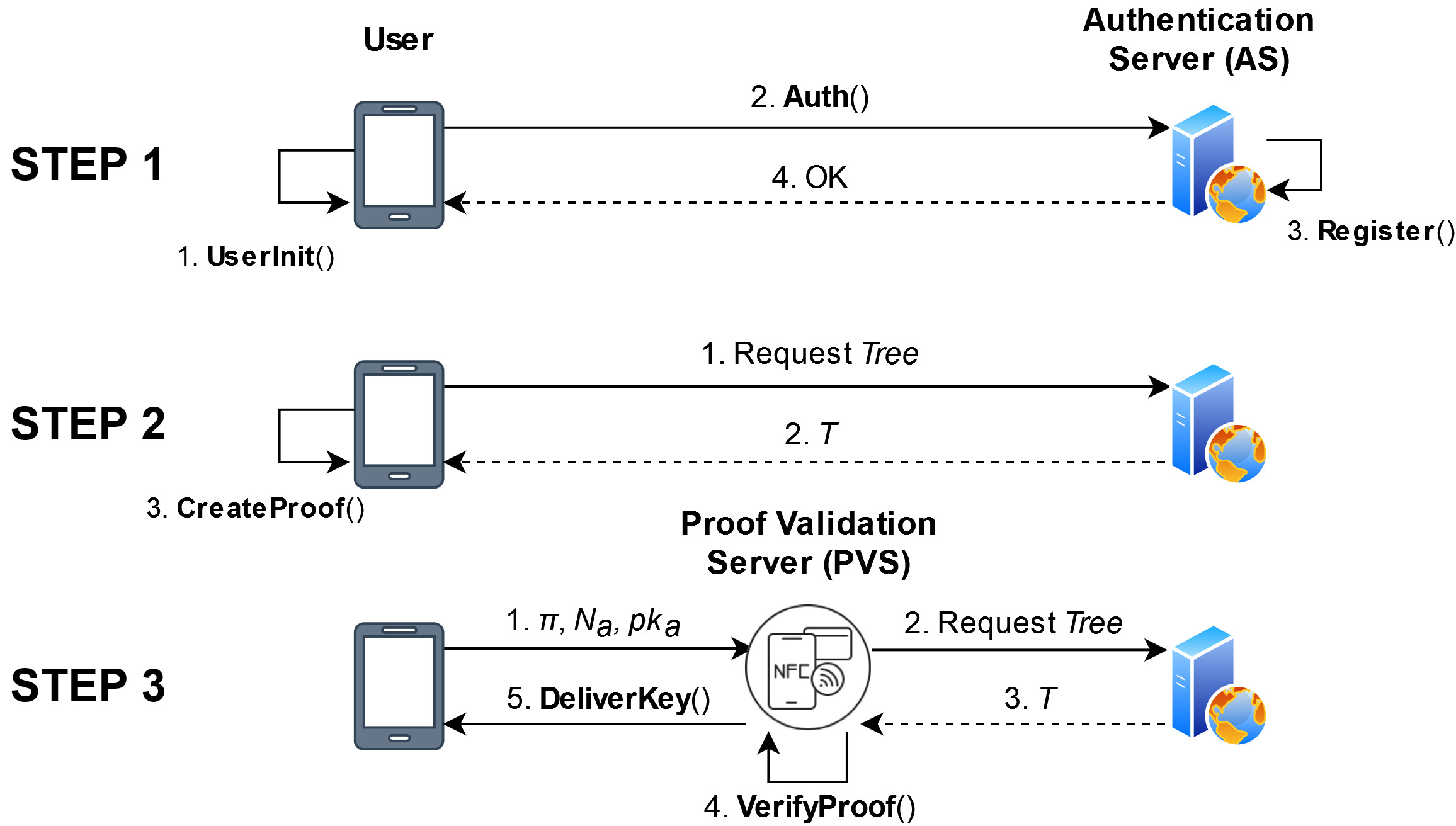}
    \caption{Diagram of the proposed protocol.}
    \label{fig:seq-protocol}
\end{figure}

Figure \ref{fig:seq-protocol} shows a diagram of the interaction between the User, Authentication Server and Proof Validation Server, as sketched in \ref{sec:outline}. 

\subsection{Protocol Definition}

The protocol requires the selection of a hash function $H$, a zk-SNARK scheme \textit{ZK = (ZK.Setup, ZK.Prove, ZK.Verify)}, a Merkle Tree scheme \textit{M = (M.EmptyTree, M.AddLeaf, M.GetRoot, M.GetIndexOf, M.ValidationList, M.IsLeafOfTree)} and a KEM \textit{K = (K.KeyGen, K.Encap, K.Decap)}. Table \ref{table:notation} introduces the notations of the elements that take part in the protocol.

\begin{table}
\centering
\begin{tabular}{ll}
\hline
Name     & Description                         \\ \hline
$crs$    & Common Reference String of zk-SNARK \\
$pk_{crs}$ & Proving Key of zk-SNARK             \\
$vk_{crs}$ & Verification Key of zk-SNARK        \\
$T$      & Commitment Merkle Tree              \\
$T_{old}$  & List of previous roots of $T$              \\
$L_N$    & List of already published Nullifiers        \\
$cert$   & User certificate             \\
$ck$    & Private key of $cert$      \\ \hline
\end{tabular}
\caption{List of notations of the proposed scheme.}
\label{table:notation}
\end{table}

\paragraph{\textbf{Initial Setup}}

Before the execution of the protocol, some parameters need to be initialised. The operation executed in this initial step is:

\begin{itemize}
    \item $\mathsf{ServerSetup}(R, n) \rightarrow ((pk_{crs}, vk_{crs}), T, T_{old}, L_N)$: The keys and data structures that will be used during the execution of the protocol are initialised. The zk-SNARK scheme is initialised from its relation $R$, and an empty Merkle Tree of depth $n$ is created.
\end{itemize}

\begin{algorithm}[t]
\caption{Initial setup operations}\label{alg:ops-init}
\begin{algorithmic}

\Function{ServerSetup}{$R, n$}
\State $crs \gets \mathsf{ZK.Setup}(R)$
\State $T \gets \mathsf{M.CreateEmptyTree}(n)$
\State $T_{old} \gets \emptyset$
\State $L_N \gets \emptyset$
\State \Return{$crs, T, T_{old}, L_N$}
\EndFunction

\end{algorithmic}
\end{algorithm}

The relation that the zk-SNARK scheme is initialised with is the following:

\begin{equation}
  \label{eq:relation}
  \renewcommand*{\arraystretch}{1.33333}
  R = \left\{
    \begin{array}{@{}l@{}}
    (x, w) = \\ 
    ((N, root), (\rho, pk, sk, C, i_C, Val))
    \end{array}
    :
    \begin{array}{@{}l@{}}
     C = H(sk \mathbin\Vert \rho), \\
     N = H(pk \mathbin\Vert \rho), \\
     \mathsf{M.IsLeafOfTree}(C, i_C, Val, root)
    \end{array}
  \right\}
\end{equation}

\begin{figure}[H]
    \centering
    \includegraphics[width=0.9\columnwidth]{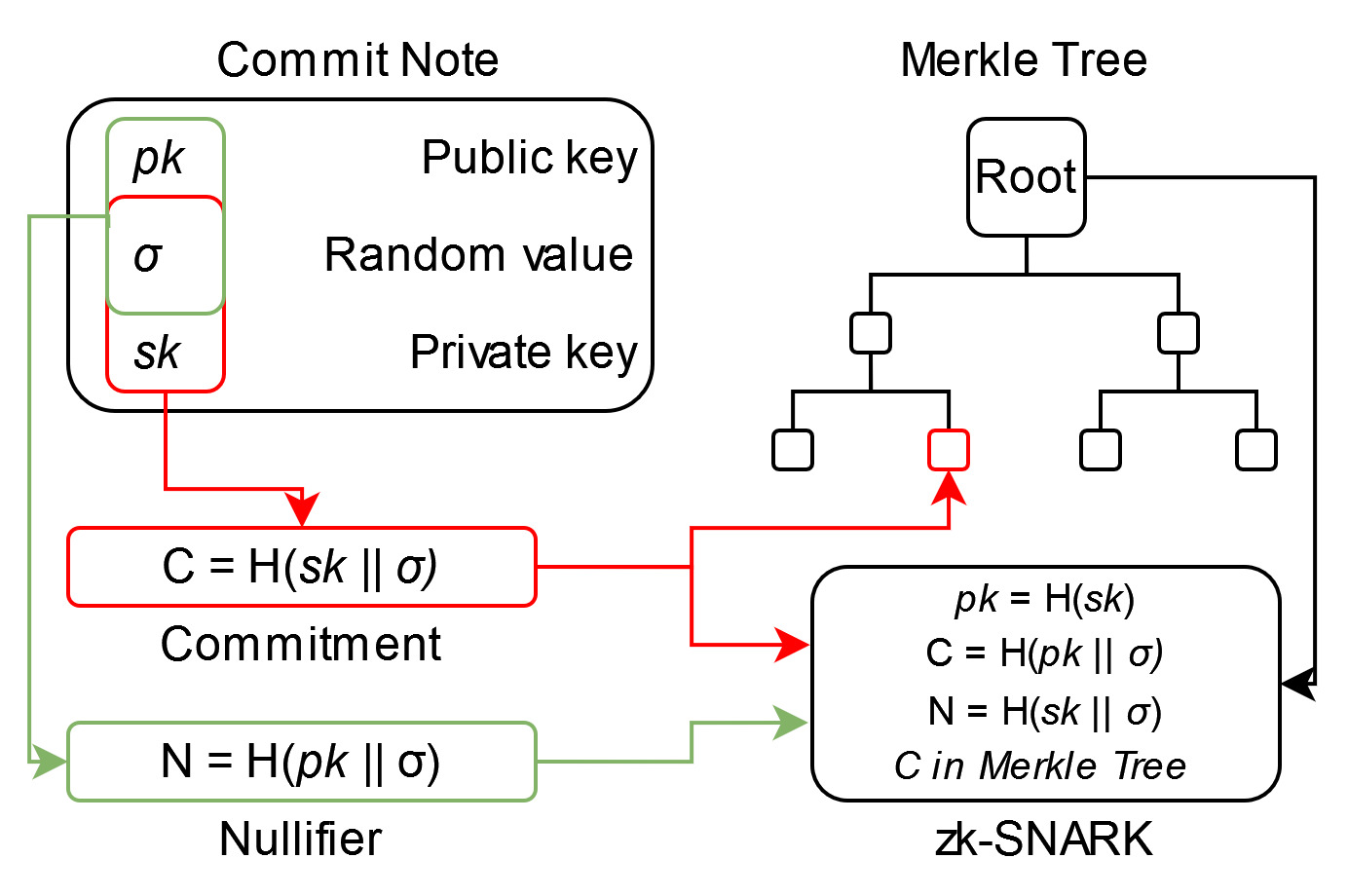}
    \caption{Diagram of the relationship between Commitments and Nullifiers.}
    \label{fig:commit}
\end{figure}

Intuitively, it proves that the Commitment and Nullifier are generated from a key pair $(pk, sk)$ and a secret value $\rho$, which is only known by the party that created both hashes. It also proves that the Commitment is inside a Merkle Tree with root $root$. This same relation is shown in Figure \ref{fig:commit}.

\paragraph{\textbf{Step 1: Authentication}}

To perform the first Step of the protocol, an User $U$ must first create a \textit{note}, which contains a secret value and a key pair. Then, she must authenticate herself to the AS. If she is successful, the AS will add $C_U$ to $T$. The Authentication Step involves the execution of the following operations (as shown in Algorithm \ref{alg:ops-auth}):

\begin{itemize}
    \item $\mathsf{UserInit}(\lambda) \rightarrow note$: $U$ creates a $note_U$ containing a secret value $\rho_U$ and an asymmetric encryption key pair $(pk_U, sk_U)$. She then stores it in a secure manner. The size of these values depends on the security parameter $\lambda$.
    \item $\mathsf{Auth}(cert, ck, note) \rightarrow (C, \sigma_C)$: $U$ uses $note_U$ to generate a Commitment $C_U = H(pk_U \mathbin\Vert \rho)$. where "$\mathbin\Vert$" is the concatenation operator. Then, she signs $C_U$ with her certificate's private key $ck_U$.
    \item $\mathsf{Register}(cert, C, \sigma_C, T, T_{old}) \rightarrow (true/false, T, T_{old})$: The AS checks the validity of the certificate $cert_U$ presented by $U$. Then, it verifies the signature $\sigma_U$ with $cert_U$'s public key. If it is valid, the AS inserts $C_U$ in the Merkle Tree $T$.
\end{itemize}

\begin{algorithm}[t]
\caption{Authentication Step operations}\label{alg:ops-auth}
\begin{algorithmic}

\Function{UserInit}{$\lambda$}
\State $\rho \overset{{\scriptscriptstyle \operatorname{R}}}{\gets} \{0,1\}^\lambda$
\State $(pk, sk) \gets \mathsf{K.KeyGen}(\lambda)$
\State $note \gets (\rho, pk, sk)$
\EndFunction

\Function{Auth}{ck, note}
\State $(\rho, pk, sk) \gets note$
\State $C \gets H(sk \mathbin\Vert \rho)$
\State $\sigma \gets \mathsf{Sign}(C, ck)$
\State \Return{$note, C, \sigma$}
\EndFunction

 \Function{Register}{$cert, C, \sigma, T, T_{old}$}
\State $b \gets \mathsf{VerifySign}(cert, C, \sigma)$
\If{$b = true$}
\State $root \gets \mathsf{M.GetRoot(T)}$
\State $T_{old} \gets T_{old} + root$
\State $\mathsf{M.AddLeaf}(T, C)$
\EndIf
\State \Return{$b$, $T$, $T_{old}$}
\EndFunction

\end{algorithmic}
\end{algorithm}

We remark that while the authentication method of the protocol we present involves the use of certificates, the Authentication Step is flexible enough to allow for other types of authentication or attestation. Different AS could be implemented that request other information from $U$ (e.g. passwords, biometrics, \ldots), as long as they insert $C_U$ in $T$ after a successful authentication.

\paragraph{\textbf{Step 2: Proof Generation}}

After a successful authentication, $U$'s Commitment $C_U$ is stored in $T$. In this Step, $U$ proves that she knows some Commitment inside $T$ by generating a zk-SNARK proof. To this end, $U$ must first download a copy of the Merkle Tree from the AS. $U$ can then use the Tree and $note_U$ to form the zk-SNARK Witness $w = (\rho_U, pk_U, sk_U, C_U, i_C, Val_U)$, which is needed to create the zk-SNARK proof. The following operation is executed, as shown in Algorithm \ref{alg:ops-gen}:

\begin{itemize}
    \item $\mathsf{CreateProof}(note, T, pk_{crs}) \rightarrow (\pi, x)$: $U$ recalculates her Commitment $C_U$ and generates the Nullifier $N_U$, both of them from the values inside $note_U$. Then, she uses the zk-SNARK scheme to create a proof $\pi_U$ that proves that $C_U$ is inside T (without revealing which element it is) and that $C_U$ and $N_U$ were generated from the same $note$.
\end{itemize}

\begin{algorithm}[t]
\caption{Proof Generation Step operations}\label{alg:ops-gen}
\begin{algorithmic}

\Function{CreateProof}{$note, T, pk_{crs}$}
\State $(\rho, pk, sk) \gets note$
\State $C \gets H(sk \mathbin\Vert \rho)$
\State $N \gets H(pk \mathbin\Vert \rho)$
\State $root \gets \mathsf{M.GetRoot}(T)$
\State $i_C \gets \mathsf{M.GetIndexOf}(T, C)$
\State $Val \gets \mathsf{M.ValidationList}(T, i_C)$
\State $x \gets (N, root)$
\State $w \gets (\rho, pk, sk, C, i_C, Val)$
\State $\pi \gets \mathsf{Prove}(pk_{crs}, x, w)$
\State \Return{$\pi, x$}
\EndFunction

\end{algorithmic}
\end{algorithm}

\paragraph{\textbf{Step 3: Key Request}}
\label{sec:step-val}

In this final Step, $U$ provides proof that she has successfully performed Step 1 to the PVS, which is the only entity in the scheme with access to the QRNG. The message from $U$ to the PVS includes $\pi_U$ and the Statement  $x = (N_U, root)$. If the information provided by $U$ is successfully validated, the PVS will reply to $U$ with key material generated by the QRNG.

The operations performed by $U$ and the PVS the Key Request Step are the following (shown in detail in Algorithm \ref{alg:ops-proof}):

\begin{itemize}
    \item $\mathsf{VerifyProof}(\pi, x, T, T_{old}, L_N, vk_{crs}) \rightarrow (true/false, L_N)$: The PVS verifies the zk-SNARK proof $\pi_U$ generated by $U$ in the previous phase. It also checks that the root $root_U$ that is part of the zk-SNARK statement $x_U$ corresponds with the root of the current Merkle Tree or any of its past versions and that the Nullifier $N_U$ has not been presented yet. If all checks are successful, $N_U$ is included in a list of "spent Nullifiers", such that it cannot be used again.
    \item $\mathsf{DeliverKey}(t, pk) \rightarrow enc$: The PVS generates $t$ bytes of key material with its QRNG and then encrypts it with $U$'s public key $pk$.
\end{itemize}

\begin{algorithm}[t]
\caption{Key Request Step operations}\label{alg:ops-proof}
\begin{algorithmic}

\Function{VerifyProof}{$\pi, x, T, L_N, T_{old}, vk_{crs}$}
\State $(N, root) \gets x$
\State $isRootValid \gets (root = \mathsf{M.GetRoot}(T) \lor root \in T_{old})$
\State $isNullifierValid \gets (N \notin L_N)$
\State $isProofValid \gets \mathsf{ZK.Verify}(vk_{crs}, \pi, x)$
\If{$isRootValid \land isNullifierValid \land isProofValid$}
\State $L_N \gets L_N + N$
\State \Return{$(true, L_N)$}
\EndIf
\State \Return{$false, L_N)$}
\EndFunction

\Function{DeliverKey}{$t, pk$}
\State $k \gets \mathsf{GenerateQRNGMaterial}(t)$
\State $enc \gets \mathsf{K.Encap}(k, pk)$
\State \Return{$enc$}
\EndFunction

\end{algorithmic}
\end{algorithm}

\section{Security} 
\label{sec:security}

In this Section we will analyse the security of our privacy-preserving key transmission protocol. We informally formulate the following security requirements that the protocol must meet:

\begin{itemize}
    \item Anonymity: it is impossible to link a User performing the Key Request Step to the credentials used in the Authentication Step. 
    \item Unforgeability: it is impossible for an attacker to successfully perform the Key Request Step without having previously performed the Authentication Step.
    \item Confidentiality: the only party that can read the QRNG key created by the PVS is the User that requested it. 
\end{itemize}

The properties of Anonymity and Unforgeability are common in protocols that employ zk-SNARKs (or other Zero Knowledge Proofs) as authentication mechanism \cite{eoltaa, elearn, nizk-auth3}. We also require the property of Confidentiality because of the inclusion of a Key Encapsulation Mechanism in our protocol. Our definition of Confidentiality is similar to the KEM's property of "One Way under Chosen Plaintext Attack" \cite{kem} but applied to the environment of our protocol.

\subsection{Anonymity} 

We will define Anonymity as a game between a Challenger $C$ and Adversary $A$. In this game, $C$ will create a pair of Commitments $(C_0, C_1)$ and choose one of them to create a Nullifier $N_b$ and proof $\pi_b$. $A$'s objective is to guess which Commitment was used to generate $(N_b, \pi_b)$. We will consider that the scheme provides Anonymity if $A$'s strategy in choosing $b$ is no better than a random guess. We remark that $A$ is allowed to execute $\mathsf{ServerSetup}$, as the Anonymity requirement also protects Users from the AS and the PVS.
 
Anonymity is defined as the following game: 
\begin{enumerate}
    \item $A$ executes $\mathsf{ServerSetup}(R, n)$ to generate $((pk_{crs}, vk_{crs}), T_0, T_{old}, L_N)$. Then, $A$ sends $(pk_{crs}, vk_{crs})$ to $C$.
    \item $C$ executes twice the following steps ($b \in \{0, 1\}$):
        \begin{enumerate}
        \item $\mathsf{UserInit}(\lambda)$ to obtain $note_b$.
        \item From $note_b$, generate the Commitment $C_b$.
        \item $\mathsf{M.AddLeaf}(T_k, C_k)$ to obtain $T_{k+1}$.
    \end{enumerate}
    \item $C$ randomly chooses $b \in \{0, 1\}$ and executes $\mathsf{CreateProof}(note_b, T_2, pk_{crs})$ to obtain $(\pi_b, (N_b, root))$. $C$ sends $((C_0, C_1), N_b, \pi_b, root)$ to $A$.
    \item $A$ outputs $b' \in \{ 0, 1 \}$.
\end{enumerate}
$A$ wins the game if $b' = b$.

We define $A$'s advantage in breaking Anonymity as 
\begin{equation}
  \label{eq:adv-anon}
    Adv^{Anon}_A(\lambda) = \abs*{\Pr [\textit{A wins the game}] - \frac{1}{2}}
\end{equation}.

\begin{definition}
The scheme provides Anonymity if for all efficient adversary $A$, $Adv^{Anon}_A(\lambda)$ is negligible.
\end{definition}

\begin{theorem}
If the hash function H is a one-way function, and the zk-SNARK scheme satisfies the Zero-Knowledge property, the proposed system provides Anonymity.
\end{theorem}

\begin{proof}
We create a series of games from Game0 to Game2.

\begin{itemize}
    \item Game0 is a real game where the adversary, $A$, tries breaking Anonymity in real attack scenarios.
    \item Game1 is the same as Game0, but the hash function $H$ is replaced by a random oracle. Since $H$ is preimage-resistant, Game1 is indistinguishable from Game0. This means that $A$ cannot obtain $note_b$ from $N_b$ or $(C_0, C_1)$.
    \item Game2 is the same as Game1, but $\pi_b$ is changed to a simulated $\pi'_b$ instead of being generated by a real $w_b$. Due to the Zero-Knowledge property of the zk-SNARK, $\pi_b$ and $\pi'_b$ are indistinguishable and the probability of $A$ winning Game 2 is the same as winning Game 1.
\end{itemize}

In Game2, none of the information available to $A$ reveals anything about $note_b$: the hash functions are not reversible and the zk-SNARK does not leak any information about its witness. Thus, $A$ cannot know if $N_b$ was generated from the same Note as $C_0$ or $C_1$ and $A$'s probability of correctly guessing $b$ is no better than a random guess. Because of this, the proposed system meets the Anonymity requirement.
\end{proof}

\subsection{Unforgeability}
We will define Unforgeability as a game between Challenger $C$ and Adversary $A$. In this game, $A$ will try to generate a proof $\pi_A$ without knowing the $note$ from which any of the Commitments in the Merkle Tree were generated. We will consider that the system provides Unforgeability if the probability of $\pi_A$ being a valid proof is negligible.

Unforgeability is defined as the following game: 
\begin{enumerate}
    \item $C$ runs $\mathsf{ServerSetup}(R, n)$ to generate $((pk_{crs}, vk_{crs}), T_0, T_{old}, L_N)$. $C$ then executes $n$ times:
    \begin{enumerate}
        \item $\mathsf{UserInit}(\lambda)$ to obtain $note_k$.
        \item From $note_k$, generate the Commitment $C_k$.
        \item $T_{old} \gets T_{old} + \mathsf{M.GetRoot}(T_k)$.
        \item $\mathsf{M.AddLeaf}(T_k, C_k)$ to obtain $T_{k+1}$.
    \end{enumerate}
    \item $C$ sends $(T_n, T_{old}, pk_{crs}, vk_{crs})$ to $A$.
    \item $A$ can execute any number of \textit{proof generation queries}, which are defined as:
    \begin{enumerate}
        \item $A$ chooses any $l \leq n$ and sends it to $C$.
        \item $C$ uses $note_l$ to execute $\mathsf{CreateProof}(note_l, T_n, pk_{crs})$ to generate $(\pi_l, (N_l, root))$. $C$ sends $(\pi_l, N_l, root)$ to $A$.
    \end{enumerate}
    \item $A$ generates $(\pi_A, N_A, root_A)$.
\end{enumerate}
$A$ wins if $\mathsf{VerifyProof}(\pi_A, (N_A, root_A), T_n, T_{old}, {N_0, ..., N_l}, vk_{crs})$ outputs $true$.

We define $A$'s advantage in breaking Unforgeability as 
\begin{equation}
  \label{eq:adv-unf}
    Adv^{Unf}_A(\lambda) = \Pr [\textit{A wins the game}]
\end{equation}.

\begin{definition}
The scheme provides Unforgeability if for all efficient adversary $A$, $Adv^{Unf}_A(\lambda)$ is negligible.
\end{definition}

\begin{theorem}
If the hash function H is a one-way function, and the zk-SNARK scheme satisfies the Knowledge Soundness and Weak Simulation-Extractability properties, the proposed system provides Unforgeability.
\end{theorem}

\begin{proof}
We create a series of games from Game0 to Game5.

\begin{itemize}
    \item Game0 is a real game where the adversary, $A$, tries breaking Unforgeability in real attack scenarios.
    \item Game1 is the same as Game0, but the hash function $H$ is replaced by a random oracle. Since $H$ is preimage-resistant, Game1 is indistinguishable from Game0. This means that $A$ cannot obtain any value inside $note_k$ from any of the $C_k$ in $T_n$.
    \item Game2 is the same as Game1, but $A$ forges a fake Merkle Tree $T_f$ which includes a fake Commitment $C_f$. Since $H$ is a random oracle, the root of $T_f$ will be different than any $T_k$'s root because their leaves are different. Thus, Condition (2) is not met and $A$ does not gain any advantage: Game2 is indistinguishable from Game1.
    \item Game3 is the same as Game2, but $A$ creates a Fake Commitment $C_f$ and forges a fake validation list $Val_f$ leading to the root of any $T_k$. From this, $A$ will generate a proof $\pi_f$. This requires finding a collision, in which two different values have the same hash (the root of $T_k$). Since $H$ is a random oracle, it will be impossible to find a collision that would falsely verify that $C_f$ is in $T_k$. Thus, the relation described in Equation~\eqref{eq:relation} will not hold and the Knowledge Soundness property of the zk-SNARK scheme will ensure that Condition (3) is not met. Thus, Game3 is indistinguishable from Game2.
    \item Game4 is the same as Game3, but $A$ is able to generate a zk-SNARK $\pi_A$ without knowing any of the values in the witness $w_A = (\rho_A, pk_A, sk_A, C_A, i_C, Val_C)$. Due to the Knowledge Soundness property of the zk-SNARK, Condition (3) will not be met with overwhelming probability. Therefore, Game4 does not increase $A$'s probability of winning over Game3.
    \item Game5 is the same as Game4, but $A$ is able to forge a fake proof $\pi_f$ by modifying a zk-SNARK $\pi_l$ received during Step 4 of the attack game (i.e. if the zk-SNARK scheme is vulnerable to malleability attacks~\cite{zk-time}). There are two possibilities in this case:
    \begin{itemize}
        \item $A$ presents $(\pi_f, N_l)$, that is, $A$ does not modify the Nullifier. In this case, $A$ fails because of Condition (1).
        \item $A$ modifies $N_l$ into a different $N_f$. Due to the Weak Simulation-Extractability property of the zk-SNARK scheme, the output of $\mathsf{Verify}$ will be $Reject$, since it is assumed to be impossible to generate a valid proof for a modified statement. Thus, $A$ fails because of Condition (3).
    \end{itemize}
    In both cases, the probability of $A$ winning Game5 is the same as winning Game4. 

\end{itemize}

In Game5, the probability achieving any of the conditions required to break the Unforgeability requirement is negligible. Since Game0 and Game5 are indistinguishable from each other, the proposed protocol meets the Unforgeability requirement. 
\end{proof}

\subsection{Confidentiality}
We will define Confidentiality as a game between Challenger $C$ and Adversary $A$. In this game, $C$ will generate a Commitment and Nullifier pair from a $note$, and an encrypted QRNG key using the public key in said $note$. We will consider that the system provides Confidentiality if $A$'s strategy in finding the QRNG key is no better than a random guess.

Confidentiality is defined as the following game: 
\begin{enumerate}
    \item $A$ executes $\mathsf{ServerSetup}(R, n)$ to generate $((pk_{crs}, vk_{crs}), T_0, T_{old}, L_N)$. Then, $A$ sends $(pk_{crs}, vk_{crs})$ to $C$.
    \item $C$ executes the following steps:
        \begin{enumerate}
        \item $\mathsf{UserInit}(\lambda)$ to obtain $note = (\rho, pk, sk)$.
        \item From $note$, generate the Commitment $C$.
        \item $\mathsf{M.AddLeaf}(T_0, C)$ to obtain $T_1$.
        \item $\mathsf{CreateProof}(note, T_1, pk_{crs})$ to obtain $(\pi, (N, root))$ 
        \item $\mathsf{DeliverKey}(t, pk)$ to generate $enc$. 
    \end{enumerate}
    \item $C$ sends $(C, N, \pi, pk, t, enc)$ to $A$.
    \item $A$ outputs $y$.
\end{enumerate}
$A$ wins if $y = K.Decap(enc, sk)$.

We define $A$'s advantage in breaking Confidentiality as 
\begin{equation}
  \label{eq:adv-conf}
    Adv^{Conf}_A(\lambda, t) = \abs*{\Pr [\textit{A wins the game}- \frac{1}{2^t}]}
\end{equation}.

\begin{definition}
The scheme provides Confidentiality if for all efficient adversary $A$, $Adv^{Conf}_A(\lambda, t)$ is negligible.
\end{definition}

\begin{theorem}
If the hash function H is a one-way function, the zk-SNARK scheme satisfies the Zero-Knowledge property and the Key Encapsulation mechanism is a one-way trapdoor function scheme, the proposed system provides Confidentiality.
\end{theorem}

\begin{proof}
We create a series of games from Game0 to Game3.

\begin{itemize}
    \item Game0 is a real game where the adversary, $A$, tries breaking Confidentiality in real attack scenarios.
    \item Game1 is the same as Game0, but the hash function $H$ is replaced by a random oracle. Since $H$ is preimage-resistant, Game1 is indistinguishable from Game0. This means that $A$ cannot obtain $sk$ from $N$ or $C$.
    \item Game2 is the same as Game1, but $\pi$ is changed to a simulated $\pi'$ instead of being generated by a real $w$. Due to the Zero-Knowledge property of the zk-SNARK, $\pi$ and $\pi'$ are indistinguishable and the probability of $A$ winning Game2 is the same as winning Game1. Since $\pi'$ was not generated by a real $w$, it is impossible to obtain $sk$ from $\pi'$. Thus, none of the values received by $A$ reveal anything about $sk$.
    \item Game3 is the same as Game2, but the Key Encapsulation Method' $\mathsf{Encap}$ function is replaced by a random oracle. Since in Game2 $A$ does not know the trapdoor $sk$, Game3 is indistinguishable from Game2.
\end{itemize}

In Game3, the value $enc$ received by $A$ is random and not related to the QRNG-generated key $y$. Since Game3 is indistinguishable from Game0, the proposed system meets the Confidentiality requirement.
\end{proof}

\section{Implementation}
\label{sec:impl}

We provide an implementation of our protocol to demonstrate its applicability. We represent a real use case in which the QRNG keys are transmitted through NFC to the User's mobile device, such that these keys could be used to securely encrypt the User's information inside their phone.
Our implementation is composed of the following elements:

\begin{itemize}
    \item Three Java programs representing the User, Authentication Server and Proof Validation Server. Both servers are located in the same machine, and have local access to their shared files. The AS is a REST API implemented with Spring Boot. Since the servers do not need to keep any information private, the AS allows Users to request any of its data structures: Merkle Tree, Nullifier List and the list of old Merkle Roots.
    \item A library tasked with the implementation of the zk-SNARK scheme, including circuit definition and proof generation and validation.
    \item A QRNG device for the generation of the key material.
    \item An NFC tunnel for the communication between the User and the PVS. To this end, we have implemented an Android App that employs the User code and acts as an NFC Smart Card. The PVS has access to an \textit{ACR 122U USB} NFC Reader to receive requests from Users.
\end{itemize}

\subsection{zk-SNARKs}

All operations related to zk-SNARK generation and validation in this work have been implemented with two different libraries: libsnark~\cite{libsnark} and the ZoKrates toolbox~\cite{zok}, which will be referred as \textit{zk-SNARK backends}. Both of these libraries require defining the statements that the zk-SNARK should prove, which in this protocol correspond to the statements that were specified in Section \ref{sec:protocol}.

Both zk-SNARK backends require a similar process for the definition of the zk-SNARK scheme. First, the relation $R$ (in this case, the relation shown in Equation \ref{eq:relation} must be codified in the zk-SNARK backend's specific language. Then, this code is compiled and an arithmetic circuit representing the calculation is generated. For both zk-SNARK backends, Users can interact with the compiled program through the command line.

Once compiled, the Common Reference String (CRS) can be generated from the arithmetic circuit. The CRS is composed of the Proving Key and the Verification Key. To execute $\mathsf{Verify}$, Users must provide both the Statement $x = (root, N)$ and the Witness $w = (\rho, pk, sk, C, i_C, Val)$. To $\mathsf{Verify}$ a proof $\pi$, only the Statement is required.

We have chosen Groth16 \cite{groth16} as the zk-SNARK scheme, since it achieves the fastest proof generation and validation times \cite{zk-time}. Moreover, it satisfies all properties mentioned in Section \ref{sec:snark}~\cite{malleable}. We remark that the proposed protocol only requires the notion of Weak Simulation-Extractability to provide Unforgeability. 

Since both zk-SNARK backends require knowing the number of parameters in compilation time, the Commitment Merkle Tree must be of fixed depth, limiting the maximum number of possible Commitments. The impact of this restriction will be studied in Section \ref{sec:test-proof}.

\subsection{QRNG}
\label{sec:qrng}

For this work, two different QRNGs have been employed:
\begin{itemize}
    \item \textit{IDQuantique's "Quantis QNRG USB"}: This device is accessible via an USB interface, and its manufacturers provide an API for multiple programming languages. Its key generation rate is about 4 Mb/s.
    \item \textit{Centro de Supercomputación de Galicia's (CESGA) QRNG module}: This supercomputer has a native QRNG infrastructure that can generate keys at 400 Mb/s. A remote SSH connection is required to gain access to this high-rate source, which introduces an overhead for establishing the connection and transmitting the key material.
\end{itemize}

\subsection{NFC Tunnel}
\label{sec:impl-nfc}

Near Field Communication (NFC) \cite{nfc1, nfc2, nfc3, nfc4} is a technology that allows wireless communication between two physically adjacent devices: an active Reader (which must always initiate the communication) and a passive Smart Card (which can only respond to the Reader's messages). NFC messages are encapsulated in Application Protocol Data Units (APDUs).

NFC is considered more secure than other wireless communication technologies such as Bluetooth because of its very small area of transmission (which prevents eavesdroppers from capturing APDUs) and its fast setup phase. Most mobile devices currently support NFC, with the capability of simulating a Smart Card and dynamically creating APDUs to respond to NFC readers. This makes NFC a very comfortable way to distribute key material to users, as it can be stored in their mobile device.

\begin{figure}[H]
    \centering
    \includegraphics[width=0.5\columnwidth]{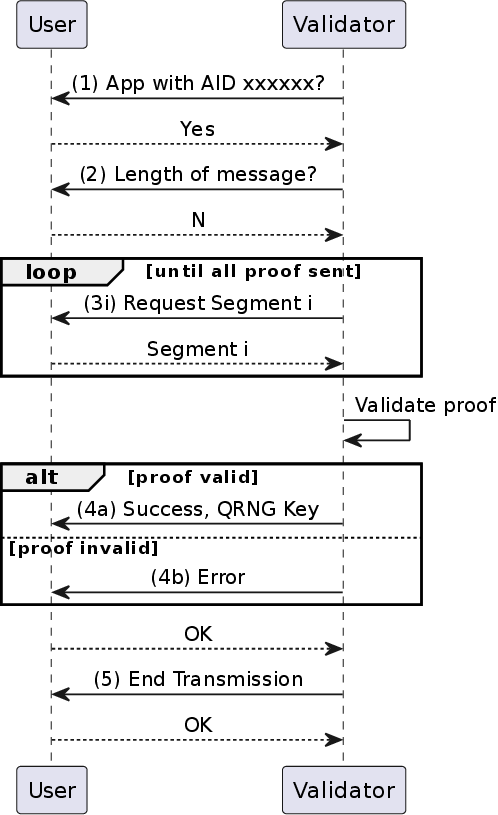}
    \caption{Sequence diagram of the NFC Tunnel.}
    \label{fig:nfc-tunnel}
\end{figure}

In our implementation, all communication between Users and the PVS in Step 3 is performed through NFC. The PVS' NFC reader interacts with $U$'s mobile device, which contains an App that can execute our protocol. Since the reader must always initiate the communication, an specific scheme is required for transmitting the zk-SNARK proof from the User to the PVS, which is shown in \ref{fig:nfc-tunnel}.

The PVS must first make sure that $U$'s mobile device contains the corresponding App. If correct, the PVS asks for the length of the message to be sent, and then performs as many requests as needed to receive the complete proof. The proof will be divided into segments to account for the small size of APDUs, which is usually limited to 255 bytes. The PVS then checks if the proof is valid and sends the QRNG if correct or an error otherwise. Finally, the PVS sends an "End Transmission" message to signal that both agents can free the employed resources.

\section{Experimental results}
\label{sec:test}

In this Section we analyse the performance of the implementation introduced in the previous Section. All of the following tests have been executed in a simulation environment, whose properties are shown in Table \ref{table:env}.

\begin{table}
\begin{tabular}{ll}
\hline
Parameter            & Value                                                                                       \\ \hline
Operating System     & Windows 10                                                                                  \\
CPU                  & Intel Core i7-10700                                                                         \\
Clock Speed          & 2.90 GHz                                                                                    \\
Memory               & 32 GB                                                                                       \\ \hline
zk-SNARK tool        & ZoKrates 0.8.3; libsnark                           \\
zk-SNARK scheme      & Groth16                                                                                     \\
Programming language & Java 17 (User, AS, PVS); ZoKrates, C++ (zk-SNARKs) \\
Hash function        & SHA-256                                                                                     \\
KEM                  & RSA-OAEP                                                                                   
\end{tabular}
\label{table:env}
\caption{Parameters of the simulation environment.}
\end{table}

\subsection{Proof Generation}
\label{sec:test-proof}

We will measure how proof generation time is influenced by the depth of the Merkle Tree. This value determines the maximum amount of Commitments it can store. However, increasing this value also increases the amount of hashes during the execution of $\mathsf{M.IsLeafOfTree}$. As mentioned in Section~\ref{sec:security}, the efficiency of proof generation is heavily dependent on the number of constraints of the arithmetic circuit. The function $\mathsf{M.IsLeafOfTree}$ is part of the relation $R$ shown in Equation~\eqref{eq:relation}, so increasing the depth of the Merkle Tree will negatively impact the proof generation time.

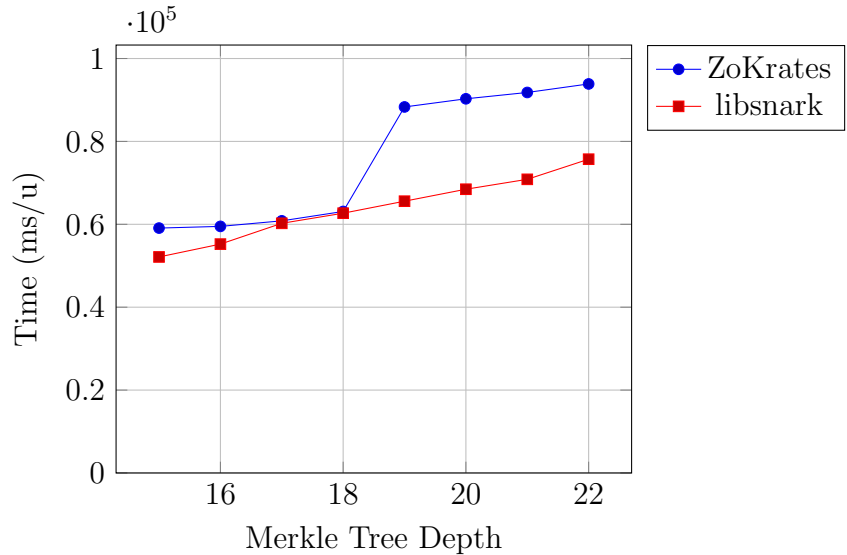
\begin{figure}[H]
    \centering
    \begin{tikzpicture}
        \begin{axis}[
        grid=major,
        legend pos=outer north east,
        xlabel={Merkle Tree Depth},
        ylabel={Time (ms/u)},
        ymin=0,
        ]
        \foreach \i in {1,...,2}{
        \addplot table [y index=\i] {data/proof_times.txt};
        }
        \legend{ZoKrates, libsnark}
        \end{axis}
    \end{tikzpicture}
    \caption{Proof generation times for different Merkle Tree depths.}
    \label{plot:test-proof}
\end{figure}

Figure \ref{plot:test-proof} shows the influence of the Merkle Tree depth proof generation time. Validation time has been omitted as it remained constant around 30 ms. The influence of the Merkle Tree depth has been tested for both zk-SNARK backends, of which libsnark achieves lower proof generation times. ZoKrates starts performing noticeably worse when Merkle Tree depth becomes 19 or higher. While Tree size (i.e., the amount of Commitments that can be stored) increases exponentially with depth, proof generation time increases only linearly.

\subsection{Key Transmission}
\label{sec:test-nfc}

The Key Request Step can be divided into four different actions: transmission of the zk-SNARK proof from $U$ to the PVS, proof validation, QRNG key generation and the transmission of the QRNG key from the PVS to $U$. Since proof size and validation time are constant according to the definition of zk-SNARKs, only QRNG key generation and transmission are variable.

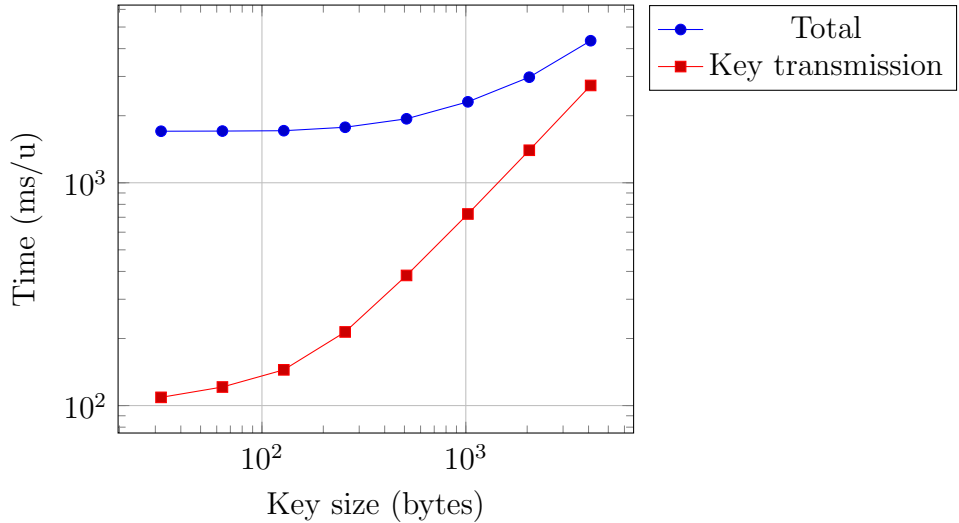
\begin{figure}[H]
    \centering
    \begin{tikzpicture}
        \begin{axis}[
        grid=major,
        xmode=log,
        ymode=log,
        legend pos=outer north east,
        xlabel={Key size (bytes)},
        ylabel={Time (ms/u)},
        ]
        \foreach \i in {1,...,2}{
        \addplot table [y index=\i] {data/transmission_times.txt};
        }
        \legend{Total, Key transmission}
        \end{axis}
    \end{tikzpicture}
    \caption{Key Request Step time, with QRNG key transmission being also shown for comparison.}
    \label{plot:test-key-send}
\end{figure}

We measured the total duration of the Key Request Step for different requested key sizes. Results are shown in Figure \ref{plot:test-key-send}: when key sizes are small, most of the time is spent transmitting the zk-SNARK and performing validation. As keys grow in size, key transmission becomes an important factor, as NFC has a relatively small transmission rate. When requesting keys of 4096 bytes in size, around 60 per cent of time is spent on transmitting the QRNG key over NFC. 

\begin{figure}[H]
    \centering
    \begin{tikzpicture}
        \begin{axis}[
        grid=major,
        xmode=log,
        ymode=log,
        legend pos=outer north east,
        xlabel={Key size (bytes)},
        ylabel={Time (ms/u)},
        ]
        \foreach \i in {1,...,4}{
        \addplot table [y index=\i] {data/gen_times.txt};
        }
        \legend{Quantis, CESGA (Total), CESGA (Gen), CESGA (SSH)}
        \end{axis}
    \end{tikzpicture}
    \caption{Key generation time for both QRNGs.}
    \label{plot:test-key-cesga}
\end{figure}
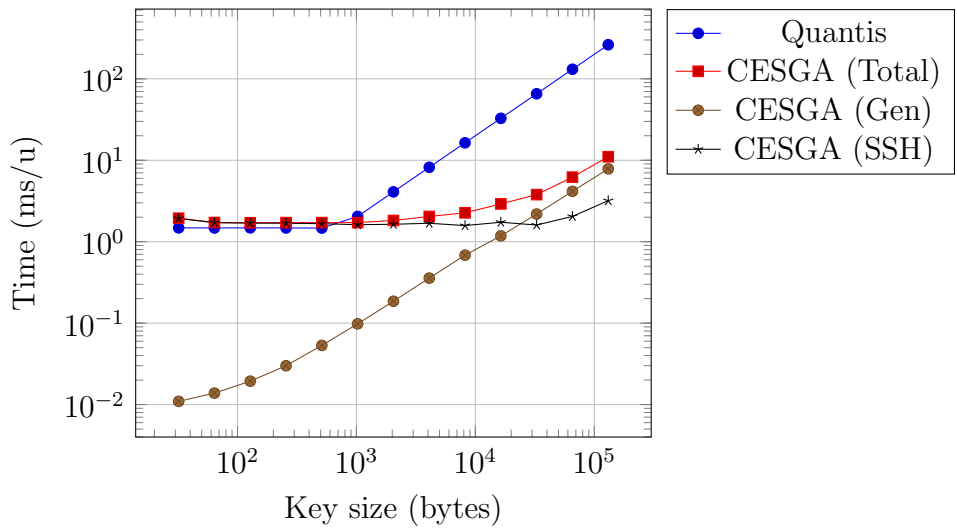

Key generation time is also influenced by requested key size. Quantis' QRNG and CESGA's QRNG are compared in Figure \ref{plot:test-key-cesga}. CESGA's time has been divided into two categories: "Generation" and "SSH". The latter corresponds to the time spent in establishing a remote connection with the QRNG and transmitting the generated key material back to the PVS. 

For small key sizes, Quantis performs better than CESGA, because most of CESGA's time is spent in establishing a remote SSH connection. CESGA's higher key generation rate becomes noticeable when requested key size increases, and SSH tunnel establishing time becomes less relevant. SSH time also includes transmitting the key from CESGA's QRNG to the PVS, so it starts to increase when requesting large amounts of key material.

We remark that zk-SNARK proof generation, which represents the most expensive computational workload of our protocol, is performed offline by the Users and thus can be executed by any number of them at the same time. The AS and PVS can handle a large amount of concurrent Users: as shown in Figures \ref{plot:test-key-send} and \ref{plot:test-key-cesga}, the time costs of the Authentication and Key Request Steps are negligible compared to that of the Proof Generation Step.

\section{Conclusion and Future Work}
\label{sec:conclusion}

This work has presented a privacy-preserving key transmission protocol for QRNG-generated cryptographic 
material, in which zk-SNARKs are employed to ensure that the server is oblivious of the authenticated 
Users requesting each key. Consequently, users do not have to trust the servers to obtain secure keys. 
The protocol could be employed to provide high-entropy key material to end users, which they could 
plug into their communications protocol stack or their applications, but we emphasize that the 
protocol is totally agnostic of the source used to generate the keys. For instance, we can 
envision our solution as a component in a QKD metropolitan network. In this architecture, 
a predefined number of nodes are connected through QKD channels such that any pair of points can 
securely establish a shared secret between them. Since access to these QKD nodes could be very
restricted, and effectively only available for high-security entities or application domains, the 
QKD keys would have to be distributed outside the quantum-enabled domain in an anonymous and 
protected form to extend the reach of QKD. To that end, an algorithmic solution like the one 
proposed in this paper can be useful. 

We provide formal security proofs to the properties of Anonymity, Unforgeability and Confidentiality.
Thus, we have proven that performing the Key Request Step does not reveal any information about which 
credentials were used in the Authentication Step, it is not possible to create a valid proof without 
first performing the Authentication Step, and only the User that performed the Key Request Step can 
access the QRNG key. Our implementation of the proposed protocol demonstrates its applicability to 
a real use case, in which the QRNG-generated keys are directly deployed to the Users' mobile phone
through NFC using a custom method for transmitting arbitrary data encapsulated in APDUs We have
implemented the zk-SNARK generation and validation operations with two different libraries and 
measure their performance under different circumstances. We plan to further develop our 
proof-of-concept Android Application to provide real functionality for the QRNG-generated key 
material.

\section*{Declaration of competing interest}

The authors declare that they have no known competing financial interests or personal relationships that could have appeared to influence the work reported in this paper.

\section*{Acknowledgement}

The work is funded by the Plan Complementario de Comunicaciones Cu\'anticas, Spanish Ministry of Science and Innovation(MICINN), Plan de Recuperación NextGeneration, European Union (PRTR-C17.I1, CITIC Ref. 305.2022), and Regional Government of Galicia (Agencia Gallega de Innovación, GAIN, CITIC Ref. 306.2022) D.S. acknowledges support from Xunta de Galicia and the European Union (European Social Fund - ESF) scholarship [ED481A-2023-219].

This work is part of the project TED2021-130369B-C31 and TED2021-130492B-C21 funded by MCIN/AEI/10.13039/501100011033 and by “ERDF A way of making Europe”.

This work has been developed thanks to the access granted by the Centro de Supercomputación de Galicia to the infrastructure based in quantum technologies. This infrastructure was financed by the European Union, through the FONDO EUROPEO DE DESENVOLVEMENTO REXIONAL (FEDER), as part of the Union's response to the COVID-19 pandemic.

\bibliographystyle{elsarticle-num} 
\bibliography{main}

\end{document}